\documentclass[10pt,a4paper]{article}

\usepackage[T1]{fontenc}
\usepackage{lmodern}
\usepackage{amsmath,amssymb,booktabs}

\newcommand\nc{\newcommand*}  \nc\longnc{\newcommand}
\longnc\VOMIT[1]{#1}  \longnc\OMIT[1]{}

\nc\subheader{\m { \qldskw{} , \qlsskw{} } in \m { \qMpos } &
  \m { \qldskw{} , \qlsskw{} } in \m { \qMmin } }

\nc\lat\textit 
\nc\ie{\lat{i.e.,\ }} \nc\etal{\lat{et al.}} \nc\etc{\lat{etc.\ }}
\nc\eg{\lat{e.g.,\ }} \nc\insitu{\lat{in situ}} \nc\QED{\lat{Q.E.D.}}
\nc\YES{\checkmark}  \nc\NO{\m { \times }}

\nc\mathBf{\boldsymbol}

\nc\m[1]{$ #1 $} 
\nc\mm[1]{\m{ \, #1 \, }} 
\nc\mmm[1]{\m{ \,\, #1 \,\, }} 
\nc\mmmm[1]{\m{ \,\,\, #1 \,\,\, }} 
\nc\M[3]{\scase=0$ 
	\mskip#1mu#3\mskip#2mu$} 
\nc\re[1]{(\ref{#1})}

\nc\0[2]{\ifcase#1{#2}\or\lt(#2\rt)\or\lt[{#2}\rt]\or\lt\{{#2}\rt\}\or
	\mathord<{#2}\mathord>\or\lt\langle{#2}\rt\rangle\or\lt\lvert{#2}\rt
	\rvert\or\lt\lVert{#2}\rt\rVert\fi}
\nc\1[2]{\ifcase#1{#2}\or(#2)\or[#2]\or\{#2\}\or\mathord<{#2}\mathord
	>\or\langle{#2}\rangle\or\lvert{#2}\rvert\or\lVert{#2}\rVert\fi}
\nc\2[2]{\ifcase#1{#2}\or\big(#2\big)\or\big[#2\big]\or\big
	\{#2\big\}\or\big<#2\big>\or\big\langle#2\big\rangle\or\big
	\lvert#2\big\rvert\or\big\lVert#2\big\rVert\fi}
\nc\3[2]{\ifcase#1{#2}\or\Big(#2\Big)\or\Big[#2\Big]\or\Big\{#2\Big
	\}\or\Big<#2\Big>\or\Big\langle#2\Big\rangle\or\Big\lvert#2\Big
	\rvert\or\Big\lVert#2\Big\rVert\fi}
\nc\4[2]{\ifcase#1{#2}\or\bigg(#2\bigg)\or\bigg[#2\bigg]\or\bigg
	\{#2\bigg\}\or\bigg<#2\bigg>\or\bigg\langle#2\bigg\rangle\or\bigg
	\lvert#2\bigg\rvert\or\bigg\lVert#2\bigg\rVert\fi}
\nc\5[2]{\ifcase#1{#2}\or\Bigg(#2\Bigg)\or\Bigg[#2\Bigg]\or\Bigg
	\{#2\Bigg\}\or\Bigg<#2\Bigg>\or\Bigg\langle#2\Bigg\rangle\or\Bigg 
	\lvert#2\Bigg\rvert\or\Bigg\lVert#2\Bigg\rVert\fi}
\nc\9[2]{\ifcase#1{#2}\or\left(#2\right)\or\left[#2\right]\or\left
	\{#2\right\}\or\left\langle{#2}\right\rangle\or\left\langle{#2}\right
	\rangle\or\left\lvert{#2}\right\rvert\or\left\lVert{#2}\right\rVert\fi}
\nc\lt{\mathopen{}\mathclose\bgroup\left} \nc\rt{\aftergroup\egroup\right}

\nc\bi\relax                     
\nc\bit{    \mskip1mu}  \nc\biT{    \mskip-1mu} 
\nc\bitt{   \mskip2mu}  \nc\biTT{   \mskip-2mu} 
\nc\bittt{  \mskip3mu}  \nc\biTTT{  \mskip-3mu} 
\nc\bitttt{ \mskip4mu}  \nc\biTTTT{ \mskip-4mu} 
\nc\bittttt{\mskip5mu}  \nc\biTTTTT{\mskip-5mu} 

\nc\f\frac
\nc\restr[2]{{\lt.#1\rt|}_{#2}} 
\nc\tr{\operatorname{tr}}
\nc\dd{\mathrm{d}} \nc\ddd{\bit\d} 
\nc\pd\partial
\nc\var\delta
\def\nablal{\lta{\nabla}}  \def\nablar{\rta{\nabla}}
\def\lta#1{{\overset{{\scriptscriptstyle \leftarrow}}{#1}}}
\def\rta#1{{\overset{{\scriptscriptstyle \rightarrow}}{#1}}}
\nc\doubledot{:}
\nc\singledot{\cdot}
\nc\dyad{\otimes}
\nc\Trans{{\rm T}}

\nc\tensor\mathbf
\nc\Tensor\mathBf

\nc\ident{\tensor 1}                  \nc\zero{\tensor 0}
\nc\identf{\tensor I}

\nc\brmin[2]{\left\{#1,#2\right\}}    \nc\brpos[2]{\left[#1,#2\right]}
\nc\rev[1]{{#1}_{\rm rev}}            \nc\irr[1]{{#1}_{\rm irr}}
\nc\sym[1]{{#1}^{\rm S}}              \nc\skw[1]{{#1}^{\rm A}}
\nc\dev[1]{{#1}^{\rm d}}              \nc\sph[1]{{#1}^{\rm s}}
\nc\devv[1]{{#1}\bit^{\rm d}}         \nc\sphh[1]{{#1}\bit^{\rm s}}
\nc\el[1]{{#1}_0}                     \nc\nel[1]{\hat{#1}}
\nc\tot[1]{{#1}}
\nc\kin[1]{{#1}_{\rm kin}}
\nc\inter[1]{{#1}_{\rm int}}
\nc\intern[1]{{#1}_{\rm int, 0}}
\nc\ther[1]{{#1}_{\rm th}}            \nc\ela[1]{{#1}_{\rm el}}
\nc\qld[1]{l_{#1}^{{\rm d}}}           \nc\qls[1]{l_{#1}^{{\rm s}}}
\nc\qldsym[1]{l^{\rm d}_{\rm S}}      \nc\qlssym[1]{l^{\rm s}_{\rm S}}
\nc\qldskw[1]{l^{\rm d}_{\rm A}}      \nc\qlsskw[1]{l^{\rm s}_{\rm A}}
\nc\qlsym{l_{\rm S}}                  \nc\qlskw{l_{\rm A}}
\nc\ql[1]{l_{#1}}
\nc\kdel[1]{\delta_{#1}}
\nc\aux{{{\rm aux}}}

\nc\Mbox[1]{\makebox[19.6em][l]{$\displaystyle #1$}}
\nc\Mboxx[1]{\makebox[14.4em][l]{$\displaystyle #1$}}
\nc\Mboxxx[1]{\makebox[17em][l]{$\displaystyle #1$}}
\nc\Mboxxxx[1]{\makebox[15em][l]{$\displaystyle #1$}}
\nc\Quad{\:\:}

\nc\dA{A}  \nc\DA{\tensor{\dA}}
\nc\dB{B}  \nc\DB{\tensor{\dB}}
\nc\qA{A}
\nc\qqAv{\qA_{\qv}}
\nc\qqAde{\qA_{\deps}}                \nc\qqAse{\qA_{\seps}}
\nc\qqAs{\qA_{\hs}}
\nc\qqAdx{\qA_{\dxi}}                 \nc\qqAsx{\qA_{\sxi}}
\nc\qB{B}
\nc\qqBv{\qB_{\qv}}
\nc\qqBde{\qB_{\deps}}                \nc\qqBse{\qB_{\seps}}
\nc\qqBs{\qB_{\hs}}
\nc\qqBdx{\qB_{\dxi}}                 \nc\qqBsx{\qB_{\sxi}}
\nc\qC{C}
\nc\young{E}
\nc\youngg{\young^\parallel}
\nc\qE{E}                             \nc\qS{S}
\nc\qET{\tilde E}                     \nc\qST{\tilde S}
\nc\qqdE[1]{E^{\rm d}_{#1}}           \nc\qqsE[1]{E^{\rm s}_{#1}}
\nc\qG{G}                             \nc\qK{K}
\nc\qMmin{\tensor L}                  \nc\qMpos{\tensor M}
\nc\QMmin{\qMmin'}                    \nc\QMpos{\qMpos'}
\nc\tMmin{\tilde\qMmin}               \nc\tMpos{\tilde\qMpos}
\nc\TMmin{\tMmin'}                    \nc\TMpos{\tMpos'}
\nc\qL[1]{L_{#1}}                     \nc\qM[1]{M_{#1}}
\nc\tL[1]{\tilde{L}_{#1}}             \nc\tM[1]{\tilde{M}_{#1}}
\nc\TL[1]{\tilde{L}_{#1}'}            \nc\TM[1]{\tilde{M}_{#1}'}
\nc\qQ{\tensor Q}
\nc\qT{T}
\nc\qV{V}

\nc\qc{c}
\nc\qe{e}
\nc\te{\tilde{e}}
\nc\qje{\tensor{j}_\qe}               \nc\qjs{\tensor{j}_\hs}
\nc\qr{r}
\nc\pdr{\pd_\qr}
\nc\qs{{\el{s}}}   \nc\hs{s}
\nc\tqs{{\el{\ts}}}   \nc\ts{\tilde{s}}   \nc\qsaux{\ts_\aux}
\nc\qt{t}
\nc\qx{\tensor x}
\nc\tx{\tilde\qx}
\nc\qv{\tensor v}
\nc\qqv{v}
\nc\qz{\tensor z}

\nc\qsig{\Tensor\sigma}
\nc\dotdsig{\dev{\dot{\qsig}}}        \nc\dotssig{\sph{\dot{\qsig}}}
\nc\qsigeld{\dev\qsig_0}              \nc\qsigels{\sph\qsig_0}
\nc\qsigneld{\dev{\hat\qsig}}         \nc\qsignels{\sph{\hat\qsig}}
\nc\qqsig{\sigma}
\nc\qqsigel{\el{\qqsig}}

\nc\eps{\varepsilon}
\nc\qeps{\Tensor\eps}
\nc\deps{\dev\qeps}                   \nc\seps{\sph\qeps}
\nc\dotdeps{\dev{\dot{\qeps}}}        \nc\dotseps{\sph{\dot{\qeps}}}
\nc\ddotdeps{\dev{\ddot{\qeps}}}      \nc\ddotseps{\sph{\ddot{\qeps}}}

\nc\qxi{\Tensor\xi}
\nc\dxi{\dev\qxi}                     \nc\sxi{\sph\qxi}
\nc\dotdxi{\devv{\dot{\qxi}}}         \nc\dotsxi{\sphh{\dot{\qxi}}}
\nc\qqxi{\xi}

\nc\qlam{\lambda}
\nc\qpis{\pi_\hs}
\nc\qrho{\varrho}
\nc\qtau{\tau}

\begin{document}

\title{Kluitenberg--Verh\'as rheology of solids in the GENERIC framework}

\author{M\'aty\'as Sz\"ucs, Tam\'as F\"ul\"op\thanks{Corresponding
author, \texttt{fulop@energia.bme.hu}.}
 \\
 \\
Department of Energy Engineering,
 \\
Faculty of Mechanical Engineering,
 \\
Budapest University of Technology and Economics,
 \\
Budapest H-1521, Hungary;
 \\
Montavid
Research Group, Budapest, Hungary
}

\maketitle

 \begin{abstract}
The internal variable methodology of nonequilibrium
thermodynamics, with a symmetric tensorial internal variable, provides
an important rheological model family for solids, the so-called
Kluitenberg--Verh\'as model family \cite{asszonyi-fulop-van:2015}. This
model family is distinguished not only from theoretical aspects but also
on experimental grounds (see \cite{asszonyi-csatar-fulop:2016} for
plastics and \cite{lin-etal:2010,matsuki:1993,matsuki:2008} for rocks).
In this article, we present and discuss how the internal variable
formulation
 of
the Kluitenberg--Verh\'as model family can be
presented in the nonequilibrium thermodynamical framework GENERIC
(General Equation for the Non-Equilibrium Reversible--Irreversible
Coupling) \cite{ottinger:2005, grmela:2010,grmela:1997,ottinger:1997},
for the benefit of both thermodynamical methodologies as well as for
promising practical applications.
 \end{abstract}

\section{Introduction}

The internal variable approach of nonequilibrium thermodynamics, with a
symmetric tensorial internal variable, provides a distinguished model
family -- the Kluitenberg--Verh\'as model family
\cite{asszonyi-fulop-van:2015} (covering the Hooke, Kelvin--Voigt,
Maxwell, Poynting--Thomson and Jeffrey models as special cases) -- for
the rheology of solids. This family is significant not only from
theoretical perspective but also for experimental applications
\cite{asszonyi-csatar-fulop:2016,lin-etal:2010,matsuki:1993,matsuki:2008}.
GENERIC (General Equation for Nonequilibrium Reversible--Irreversible
Coupling) is an attractive general framework for nonequilibrium
thermodynamical models (see, \eg \cite{ottinger:2005,
grmela:2010,grmela:1997,ottinger:1997}). Whenever a new nonequilibrium
thermodynamical model emerges, it is advantageous and recommended to
check how it suits the frame of GENERIC. Here, we investigate how the
internal variable formulation leading to the Kluitenberg--Verh\'as model
family can be represented in GENERIC.

For the main part of the paper, specific entropy is treated as one of
the state variables---a choice natural from principal aspects. Later, in
an alternative version, temperature is used, instead---which formulation
may be more convenient for certain engineering applications.

We believe that the relationship between the internal variable framework
and GENERIC may be fruitful for both approaches, providing
 \begin{itemize}
 \item
insight concerning the theoretical side,
 \item
wider applicability,
 \item
suggestions for novel numerical methods (see, \eg
\cite{ottinger:2018,shang-ottinger:2018} for such a promising
direction), and
 \item
beneficial connection of such numerical approaches with analytical
results (\eg \cite{fulop-szucs:2018}).
 \end{itemize}
In this respect, this paper intends to serve as a case study.

Notably, certain aspects of our treatment could be presented at some
more general, systematic and methodological level (for example,
performing the deviatoric--spherical separation of tensors in a
multiplicative split form \cite{edwards:1998}). Here, we follow a simple
direct approach.

\section{Necessary elements I: The internal variable formulation of
rheology of solids}  \label{intsum}

We start with a summary and generalization of the internal
variable approach
to the Kluitenberg--Verh\'as rheological model family of solids
\cite{asszonyi-fulop-van:2015}. The discussion is generalized in that
the derivation in \cite{asszonyi-fulop-van:2015} neglected thermal
expansion and started from Hookean elasticity, while the version here is
free of those restrictions, only isotropy of the material being assumed.

The small-strain regime is considered, where strain \m { \qeps } is
small (\m{ \16{\qeps}\ll 1 }), there is no need to distinguish spacetime
and material manifold variables and vectors/tensors---accordingly,
aspects of objectivity and spacetime compatibility \cite{fulop-van:2012,fulop:2015} are
not addressed here---, mass density \m{\qrho} is constant, and one can
relate time derivative (partial coinciding with substantial) of strain
with the symmetrized gradient%
 \footnote{\m{\nablal} and \m{\nablar} act to the left and to the right,
 respectively, reflecting proper tensorial order.}
of the velocity field \m { \qv },
 \begin{align}  \label{kinematiceq}
\dot{\qeps} = \sym{\91{\qv\otimes\nablal}} .
 \end{align}
Due to the isotropy of the material, the deviatoric--spherical
decomposition of symmetric tensors plays here an important role (the
spherical part of, \eg strain is proportional to the identity tensor,
\m{ \seps=\f{1}{3}\91{\tr\qeps}\ident }, while \m{\deps=\qeps-\seps} is
its deviatoric part). With \m { \qs } denoting mass-specific entropy,
our variables will be
 \begin{align}  \label{statevar}
\el\qx = \91{\qv,\deps,\seps,\qs} .
 \end{align}
The balance of linear momentum is
 \begin{align}  \label{mombalance}
\qrho \dot{\qv} = \el\qsig \cdot \nablal
 \end{align}
with the divergence of stress \m { \el\qsig } on the rhs, where
\m { \el\qsig }
is related to a (mass-)specific internal energy \m { \intern{\qe} \9 1 {
\deps,\seps,\qs } } as shown by the partial derivatives\footnote{\m {
\f{\pd \intern{\qe}}{\pd \deps} } is purely deviatoric since, for
 any two
tensors \m { \DA, \DB },
\mm { \1 0 {\sph\dA}_{ij} \1 0 {\dev\dB}_{ij} = 0 }
 so \mm { \f{\pd \intern{\qe}}{\pd \dev\eps_{ij}} \dd
\dev\eps_{ij} = \dev{\3 1 { \f{\pd \intern{\qe}}{\pd \dev\eps_{ij}}
}\biT}
\dd \dev\eps_{ij} \,.} }
 \begin{align}  \label{partialderivatives}
\f{\pd \intern{\qe}}{\pd \deps}=\f{1}{\qrho}\qsigeld,
 \qquad
\f{\pd \intern{\qe}}{\pd \seps}=\f{1}{\qrho}\qsigels,
 \qquad
\f{\pd \intern{\qe}}{\pd \qs}=\qT ,
 \end{align}
\m { \qT } standing for temperature.

Rheology is a behaviour most manifest in the mechanical aspect so, for
an internal variable description of it, in conform with that strain and
stress are second order symmetric tensors, we introduce a symmetric
tensorial internal variable \m{\qxi}. Mechanical effects of rheology are
to be embodied by a \m { \qxi } dependent extension of stress:
 \begin{align}  \label{extstress}
\qsig & = \el{\qsig} + \nel{\qsig} ,
 \\  \label{mombalancee}
\qrho\dot{\qv} & = \qsig\cdot\nablal .
 \end{align}
We conceive rheology as irreversibility-related so specific entropy is
also assumed to be influenced; concavity concerns combined with Morse's
lemma for smooth enough \m { \qxi } dependence and nonzero second
derivative in \m { \qxi } conclude in the variable
transformation \m { \qs \to \hs } \cite{verhas:1997}
 \begin{align}  \label{extentropy}
\hs=\qs-\f{1}{2}\tr{\91{\dxi\dxi}}-\f{1}{2}\tr{\31{\sxi\sxi}} .
 \end{align}
Correspondingly, specific internal energy expressed in terms of the
extended collection of variables,
 \begin{align}  \label{statevarr}
\qx=\91{\qv,\deps,\seps,\hs,\dxi,\sxi} ,
 \end{align}
is of the form
 \begin{align}  \label{totenergyy}
\inter{\qe}\91{\qx}=\intern{\qe}\31{\deps,\seps,\qs\21{\hs,\dxi,\sxi}} .
 \end{align}

The balance of internal energy is
 \begin{align}
\label{energybalance}
\qrho\inter{\dot{\qe}}=-\qje\cdot\nablal+\tr\91{\qsig\dot{\qeps}},
 \end{align}
where \m{\qje} denotes heat current density, and the only source term
considered is related to mechanical power. Substituting
\re{partialderivatives}, \re{extstress} and \re{extentropy} into
\re{energybalance}, on the one hand we obtain
 \begin{align}  \nonumber
\qrho\inter{\dot{\qe}} & = \qrho \93{ \tr \91{
\f{\pd \intern{\qe}}{\pd \deps} \dotdeps } + \tr\91{
\f{\pd \intern{\qe}}{\pd \seps} \dotseps } +
\f{\pd \intern{\qe}}{\pd \qs} \92{ \f{\pd \qs}{\pd \hs} \dot{\hs}
+ \tr \91{\f{\pd \qs}{\pd \dxi} \dotdxi } +
\tr \91{\f{\pd \qs}{\pd \sxi} \dotsxi}} } =
 \\  \label{energybalance1}
&=\tr\91{\qsigeld\dotdeps}+\tr\31{\qsigels\dotseps}+
\qrho\qT\dot{\hs}+\qrho\qT\tr\91{\dxi\dotdxi}+\qrho\qT\tr\31{\sxi\dotsxi} ,
 \end{align}	
and on the other hand
 \begin{align}  \label{energybalance2}
\qrho\inter{\dot{\qe}}=-\qje\cdot\nablal+\tr\91{\qsigeld\dotdeps}+
\tr\31{\qsigels\dotseps}+\tr\91{\qsigneld\dotdeps}+\tr\31{\qsignels\dotseps}.
 \end{align}
The rhs of \re{energybalance1} is to be equal to the rhs of
\re{energybalance2},
which leads to
 \begin{align}  \label{rhodots}
\qrho\dot{\hs}=-\f{1}{\qT}\qje\cdot\nablal+
\f{1}{\qT}\tr\91{\qsigneld\dotdeps}+
\f{1}{\qT}\tr\31{\qsignels\dotseps}-
\qrho\tr\91{\dxi\dotdxi}-\qrho\tr\31{\sxi\dotsxi}.
 \end{align}
Since the balance of the extended entropy is to be of the form
 \begin{align}  \label{entropybalance}
\qrho\dot{\hs}=-\qjs\cdot\nablal+\qpis
 \end{align}
with entropy current density \m{\qjs} chosen to be the usual
\mm{\qjs=\f{1}{T}\qje}, and entropy production \m{\qpis},
in the light of \re{rhodots}, we can write
 \begin{align}  \label{entropyproduction}
\qpis=\qrho\dot{\hs}+\qjs\cdot\nablal = \qje \cdot \91{ \f{1}{\qT}
\otimes \nablal } + \f{1}{\qT} \tr \31{\qsigneld\dotdeps} +
\f{1}{\qT} \tr \31{\qsignels\dotseps} - \qrho\tr\91{\dxi\dotdxi}
-\qrho \tr \31{\sxi\dotsxi}.
 \end{align}
Positive semidefiniteness of entropy production can be ensured for the
first term via \mmm{\qje=\qlam\91{\f{1}{\qT}\otimes\nablal} \,,}
\m{\qlam \ge 0} (Fourier heat conduction, a vectorial part that cannot
isotropically couple to the remaining, tensorial, terms; hence, for
simplicity, heat conduction is neglected in what follows), and via
Onsagerian equations concerning the further terms, with independent
deviatoric and spherical parts because of isotropy:
 \begin{align}  \label{ons1}
&&&&
\qsigneld & = \qld{11} \dotdeps + \qld{12} \01{         - \qrho\qT\dxi} ,
 &
\qsignels & = \qls{11} \dotseps + \qls{12} \31{ \mathord- \qrho\qT\sxi} ,
&&&&
 \\  \label{ons4}
&&&&
\dotdxi & = \qld{21} \dotdeps + \qld{22} \01{         - \qrho\qT\dxi} ,
 &
\dotsxi & = \qls{21} \dotseps + \qls{22} \31{ \mathord- \qrho\qT\sxi} ,
&&&&
 \end{align}
with appropriate conditions on the deviatoric coefficients \m{\qld{ij}}
and the spherical ones \m{\qls{ij}}, each of which are going to be assumed
constant for simplicity\footnote{No principal difficulties are expected
when these coefficients are \m { \qx } dependent, and here we intend
to keep formulae relatively short.}. These conditions can be read off
from the quadratic form obtained by substituting \re{ons1}--\re{ons4} into
\re{entropyproduction}, which yields\footnote{The upper right and
lower left two-by-two submatrices contain only zero elements due to the
isotropic decoupling of deviatoric and spherical parts.}
 \begin{align}
\qT\qpis=
\begin{pmatrix}
\dotdeps&-\qrho\qT\dxi&\dotseps&-\qrho\qT\sxi
\end{pmatrix}
\begin{pmatrix}
\qld{11}&\qldsym{12}&0&0\\
\qldsym{12}&\qld{22}&0&0\\
0&0&\qls{11}&\qlssym{12}\\
0&0&\qlssym{12}&\qls{22}
\end{pmatrix}
\begin{pmatrix}
\dotdeps\\-\qrho\qT\dxi\\ \dotseps\\-\qrho\qT\sxi
\end{pmatrix}
 \end{align}
with \m{\qldsym{12}=\f{1}{2}\91{\qld{12}+\qld{21}}} and
\m{\qlssym{12}=\f{1}{2}\91{\qls{12}+\qls{21}}}. Hence, the four-by-four
coefficient matrix
in the middle is required to be positive semidefinite, which necessitates
for the coefficients, using Sylvester's criteria,
 \begin{align}  \label{requird}
&&&&
\qld{11} & \ge 0,
 &
\qld{22} & \ge 0,
 &
\det{\qldsym{}} &\ge 0,
&&&&
 \\  \label{requirs}
&&&&
\qls{11} &\ge 0,
 &
\qls{22} & \ge 0,
 &
\det{\qlssym{}} & \ge 0.
&&&&
 \end{align}
We remark that, both in \re{requird} and \re{requirs}, the three
conditions are not independent: the third one and either of the first
two ones imply the remaining one. It is important to notice that the
antisymmetric part of the coefficient matrix does not contribute to
entropy production.
We can emphasize this by dividing the Onsagerian equations
\re{ons1}--\re{ons4} into
two parts:
 \begin{align}  \label{ons1sa}
\qsigneld & = \92{ \qldskw{12} \91{-\qrho\qT\dxi}} +
\92{ \qld{11} \dotdeps + \qldsym{12} \01{         - \qrho\qT\dxi} },
 &
\qsignels & = \32{ \qlsskw{12} \31{-\qrho\qT\sxi} } +
\32{ \qls{11} \dotseps + \qlssym{12} \31{ \mathord- \qrho\qT\sxi} },
 \\  \label{ons4sa}
\dotdxi & = \92{ -\qldskw{12} \dotdeps } +
\92{ \qldsym{12} \dotdeps + \qld{22} \91{         - \qrho\qT\dxi} },
 &
\dotsxi & = \32{ -\qlsskw{12}\dotseps } +
\32{ \qlssym{12} \dotseps + \qls{22} \31{ \mathord- \qrho\qT\sxi} }
 \end{align}
with \mm{\qldskw{12}=\f{1}{2} \01{\qld{12}-\qld{21}}} and
\mm{\qlsskw{12}=\f{1}{2} \31{\qls{12}-\qls{21}}}.

It is to be noted that, in general, the coefficient matrices \m { \qld{}
}, \m { \qls{} } need not be symmetric nor antisymmetric, corresponding
to that the concrete physical interpretation of \m { \qxi } may not
be available and the behaviour of \m { \qxi } under time reflection
might not be purely sign preserving/flipping.\footnote{Rocks are one
example of complex enough materials that may require such a description.}

To see that this model family covers classic rheological models like
Kelvin--Voigt and Poynting--Thomson, one can start with the special case
of Hooke elasticity, and eliminating the internal variable leads, in the
isothermal approximation (constant \m { \ql{11}^{\rm d,s} },
\m { \qlskw^{\rm d,s} }, \m { \qrho\qT \qlsym^{\rm d,s} },
\m { \qrho\qT \ql{22}^{\rm d,s} }),
to the Kluitenberg--Verh\'as model family
\cite{asszonyi-fulop-van:2015},
 \begin{align}
\dev{\qsig}+\dev{\qtau}\dotdsig&=\qqdE{0}\deps+\qqdE{1}\dotdeps+\qqdE{2}\ddotdeps ,
 \\
\sph{\qsig}+\sph{\qtau}\dotssig&=\qqsE{0}\seps+\qqsE{1}\dotseps+\qqsE{2}\ddotseps ,
 \end{align}
with necessary and sufficient thermodynamical inequality conditions on
the coefficients \m{\qtau^{\rm d,s},E_{0}^{\rm d,s},E_{1}^{\rm
d,s},E_{2}^{\rm d,s}}
stemming from \re{requird}--\re{requirs} (for further details on
the elimination and the inequalities, see
\cite{asszonyi-fulop-van:2015}, Section~2.3).

\section{Necessary elements II: Summary of the GENERIC framework}

In  GENERIC \cite{ottinger:2005, grmela:2010}, time evolution of the
collection of state variables (fields, in case of continuum models like
ours here), \m{\qx}, is formulated as
 \begin{align}  \label{geneq}
\f{\dd \qx}{\dd \qt}=\qMmin\01{\qx}\f{\var\qE}{\var\qx}
+ \qMpos\01{\qx}\f{\var\qS}{\var\qx}
,
 \end{align}
where the
operator matrix
\m{\qMmin} acts on the column vector that is the functional derivative
of the energy functional \m { \qE } of \m { \qx }, and the
operator matrix \m{\qMpos} acts on the column vector that is the
functional derivative of the entropy functional \m { \qS } of \m { \qx
}. \m{\qMmin} is required to be antisymmetric,
 \begin{align}  \label{antisym}
\qMmin=-\qMmin^\Trans
 \end{align}
(\m{^\Trans}
denoting transpose which, for operators, means not merely matrix index
transposition but includes operator adjoint). Thanks to this and the
degeneracy condition
 \begin{align}  \label{crit1}
\qMpos\f{\var \qE}{\var \qx}= \zero ,
 \end{align}
energy is conserved, \mm { \f{\dd \qE}{\dd \qt} = 0 }. In parallel, the
other degeneracy requirement
 \begin{align}  \label{crit2}
\qMmin\f{\var \qS}{\var \qx}= \zero
 \end{align}
ensures that the first term on the rhs of \re{geneq} does not increase
entropy, and \m{\qMpos} is demanded to be positive semidefinite to lead
to \mm { \f{\dd \qS}{\dd \qt} \ge 0 } eventually. That \m { \qMmin
\f{\var\qE}{\var\qx} } is related to reversible dynamics is manifested
further by also prescribing the Jacobi identity
 \begin{align}
\brmin{\qA}{\brmin{\qB}{\qC}}+\brmin{\qB}{\brmin{\qC}{\qA}}+\brmin{\qC}{\brmin{\qA}{\qB}}=0
 \end{align}
(\m{\qA}, \m{\qB}, \m{\qC} arbitrary functionals)
for the bilinear generalized Poisson bracket
 \begin{align}  \label{brmin}
\brmin{\qA}{\qB} := \int_\qV \f{\var \qA}{\var \qx} \qMmin
\f{\var\qB}{\var \qx} \dd\qV .
 \end{align}
Consequently, the first term on the rhs of \re{geneq} can be interpreted
as a reversible---generalized Hamiltonian---time evolution contribution
(vector field) while the second term (another vector field) embodies
irreversible time evolution contribution to dynamics.

Analogously to \re{antisym} and \re{brmin}, imposing symmetricity for \m
{ \qMpos },
 \begin{align}  \label{sym}
\qMpos=\qMpos^\Trans ,
 \end{align}
induces that the bilinear product
 \begin{align}  \label{brpos}
\brpos{\qA}{\qB} := \int_\qV \f{\var \qA}{\var \qx} \qMpos
\f{\var\qB}{\var \qx} \dd\qV
 \end{align}
is positive semidefinite, \m{\brpos{A}{A}\ge 0}. The latter bracket
\re{brpos} completes the former one \re{brmin} in the sense that time
evolution for any functional \m { \qA } can be expressed as
 \begin{align}
\f{\dd \qA}{\dd \qt}=\brmin{\qA}{\qE}+\brpos{\qA}{\qS}.
 \end{align}

A constructive and productive way to generate the irreversible
contribution to dynamics is to derive it from a dissipation potential
\cite{grmela:2010,janecka-pavelka:2018,grmela-etal:2018}. Assuming a
dissipation potential is, on the other side, not necessary and reduces
the level of generality of the GENERIC framework
\cite{hutter-svendsen:2013}.

\section{Internal variable rheology of solids realized in the GENERIC
formulation}  \label{intgen}

Section~\ref{intsum} has actually been given in a form to provide
preparations for the present one, where we establish GENERIC form for
the \m { \qxi }-described rheology of solids. The set of variables \m {
\qx } is \re{statevarr}, the energy functional consists of the internal
energy contribution \re{totenergyy} supplemented by the kinetic energy
related one, and the entropy functional is straightforward:
 \begin{align}
\qE & = \int_\qV \qrho\tot{\qe} \dd\qV = \int_\qV \qrho
\02{ \f{1}{2} \qv\cdot\qv + \intern{\qe} \31{ \deps, \seps,
\qs\21{\hs, \dxi, \sxi} } } \dd \qV,
 \\
\qS & = \int_\qV \qrho\hs \dd\qV .
 \end{align}
The corresponding functional derivatives
are
 \begin{align}  \label{funcd}
\f{\var \qE}{\var \qx} = \begin{pmatrix}
\qrho\qv\\ \qsigeld\\ \qsigels\\ \qrho\qT\\ \qrho\qT\dxi\\ \qrho\qT\sxi
\end{pmatrix},
 \qquad\qquad
\f{\var \qS}{\var \qx} = \begin{pmatrix}
\zero\\ \zero\\ \zero\\ \qrho\\ \zero\\ \zero  \end{pmatrix} .
 \end{align}

The nontrivial task is to identify \m { \qMmin } and \m { \qMpos }.
Concerning the time evolution of the state variables, we know
\re{mombalancee}, \re{kinematiceq}, \re{rhodots} and
\re{ons1}--\re{ons4} so we conjecture the decomposition to reversible
and irreversible parts as%
 \footnote{Beware that, if we prefer to write \m { \qx } and \m {
\dot\qx } as column vectors then \re{funcd} should contain row vectors,
as being covectors with respect to the vector space of \m { \qx }.
However, then \m { \qMmin } and \m { \qMpos } could not be displayed as
customary square matrices. The misleading double meaning of column
vectors could be resolved by writing \m { \qx }-covectors like in
\re{funcd} as column vectors but within \m { [ \Quad ] } instead of \m {
( \Quad ) }. Then, correspondingly, \m { \qMmin }, \m { \qMpos } were to
be written within \m { ( \Quad ] }. Here, we decided not to use this
convention but at least to draw attention to that distinction
between vectors and covectors is not only principally important but
also avoids considerable confusion during calculations.
 }
 \begin{align}  \label{timeevolution}
 \begin{pmatrix}
\dot\qv\\\dotdeps\\\dotseps\\\dot\hs\\\dotdxi\\\dotsxi
 \end{pmatrix} = \begin{pmatrix}
\f{1}{\qrho}\02{\qsigeld+\qsigels}\cdot\nablal \\
\dev{\02{\sym{\01{\qv\otimes\nablal}}}} \\
\sph{\02{\sym{\01{\qv\otimes\nablal}}}} \\ 0 \\ \zero \\ \zero
 \end{pmatrix}
+
 \begin{pmatrix}
\f{1}{\qrho}\02{\qld{11}\dotdeps+\qld{12}\01{-\qrho\qT\dxi} +
\qls{11}\dotseps + \qls{12} \31{\mathord- \qrho\qT\sxi}}\cdot\nablal\\
\zero \\ \zero \\
 \begin{pmatrix}
\f{\qld{11}}{\qrho\qT} \tr\01{\dotdeps\dotdeps} -
2 \qldsym{12}
\tr\01{\dxi\dotdeps}+\qld{22}\qrho\qT\tr\01{\dxi\dxi} \mathrel+
 \\
\mathrel+ \f{\qls{11}}{\qrho\qT} \tr\31{\dotseps\dotseps} -
2 \qlssym{12}
\tr\31{\sxi\dotseps}+\qls{22}\qrho\qT\tr\31{\sxi\sxi}\end{pmatrix} \\
\qld{21}\dotdeps+\qld{22}\91{         - \qrho\qT\dxi} \\
\qls{21}\dotseps+\qls{22}\31{ \mathord- \qrho\qT\sxi}
 \end{pmatrix} .
 \end{align}
The governing principle for this decision for decomposition is that,
since dissipation and irreversibility are related to entropy production
and to the internal variable, the reversible vector field should not
contain them but only pure fluid mechanics.

Then \m{\qMmin} can directly read off from
the first term on the rhs of \re{timeevolution}:
 \begin{align}  \label{revop}
\qMmin =
 \begin{pmatrix}
0&\f{1}{\qrho}\bullet\cdot\nablal&\f{1}{\qrho}\bullet\cdot\nablal&\zero&\zero\singledot&\zero\singledot\\
\f{1}{\qrho}\dev{\02{\sym{\01{\bullet\otimes\nablal}}}}&0&0&\zero&0&0\\
\f{1}{\qrho}\sph{\02{\sym{\01{\bullet\otimes\nablal}}}}&0&0&\zero&0&0\\
\zero\singledot&\zero\doubledot&\zero\doubledot&0&\zero\doubledot&\zero\doubledot\\
\zero\dyad&0&0&\zero&0&0 \\ \zero\dyad&0&0&\zero&0&0
 \end{pmatrix} ,
 \end{align}
with \m { \bullet } denoting the `slot' where the operator acts. This
\m{\qMmin} apparently fulfils the degeneracy
condition \mm{\qMmin\f{\var\qS}{\var\qx}=\zero}.

To prove antisymmetry of \m{\qMmin}, let us take
the corresponding bracket \re{brmin}:
 \begin{align}
\brmin{\qA}{\qB} = \int_\qV \f{1}{\qrho} \52{ \qqAv \cdot \01{
\qqBde\cdot\nablal + \qqBse\cdot\nablal} + \tr \93{ \qqAde
\dev{ \02{ \sym{\01{\qqBv\otimes\nablal}} } } } + \tr \93{
\qqAse \sph{ \02{ \sym{\01{\qqBv\otimes\nablal}} } } } } \dd\qV ,
 \end{align}
where \m{\qA} and \m{\qB} are arbitrary functionals of the state
variables, and abbreviations of the kind
 \begin{align}
\qqAv := \f{\var \qA}{\var \qv} ,
 \quad
\qqAde := \f{\var \qA}{\var \deps} ,
 \quad
\qqAse := \f{\var \qA}{\var \seps} ,
 \quad
\qqAs := \f{\var \qA}{\var \hs} ,
 \quad
\qqAdx := \f{\var \qA}{\var \dxi} ,
 \quad
\qqAsx := \f{\var \qA}{\var \sxi}
 \end{align}
have been introduced. Using indices (with Einstein convention and the
Kronecker symbol \m{\kdel{ij}}), we have
 \begin{align}
\sph{\02{\sym{\01{\qqBv\otimes\nablal}}}} & =
\f{1}{3}\tr\02{\sym{\01{\qqBv\otimes\nablal}}}\ident
= \f{1}{3}\pd_k\01{\qqBv}_k\kdel{ij} ,
 \\
\dev{\02{\sym{\01{\qqBv\otimes\nablal}}}} & =
\sym{\01{\qqBv\otimes\nablal}} -
\sph{\02{\sym{\01{\qqBv\otimes\nablal}}}} =
\f{1}{2}\02{\pd_j\01{\qqBv}_i + \pd_i\01{\qqBv}_j} -
\f{1}{3}\pd_k\01{\qqBv}_k\kdel{ij} ,
 \end{align}
leading to
 \begin{align}
\brmin{\qA}{\qB} & = \int_\qV \f{1}{\qrho}
\Bigg[\01{\qqAv}_i \02{\pd_j\01{\qqBde}_{ij} + \pd_j\01{\qqBse}_{ij}}
+ \01{\qqAde}_{ij} \93{\f{1}{2} \02{\pd_j\01{\qqBv}_i +
\pd_i\01{\qqBv}_j} - \f{1}{3} \pd_k\01{\qqBv}_k\kdel{ij}}
\mathrel+
 \nonumber  \\  \label{pzj}
 & \hskip 4em
+ \01{\qqAse}_{ij}\f{1}{3}\pd_k\01{\qqBv}_k\kdel{ij}\Bigg]\dd\qV.
 \end{align}
Via integration by parts and omitting surface terms,
 \begin{align}
\brmin{\qA}{\qB} & = -\int_\qV \f{1}{\qrho}
\Bigg[\pd_j\01{\qqAv}_i\01{\qqBde}_{ij} + \pd_j\01{\qqAv}_i
\01{\qqBse}_{ij} +\f{1}{2} \02{\pd_j\01{\qqAde}_{ij} \01{\qqBv}_i +
\pd_i\01{\qqAde}_{ij} \01{\qqBv}_j}
\mathrel-
 \nonumber  \\
 & \hskip 5em
-\f{1}{3}\pd_k\01{\qqAde}_{ij}\01{\qqBv}_k\kdel{ij}+\f{1}{3}\pd_k\01{\qqAse}_{ij}\01{\qqBv}_k\kdel{ij}\Bigg]\dd\qV.
 \end{align}
Now, taking into consideration that the functional derivative of a
scalar functional with respect to a symmetric tensor is symmetric, with
respect to a deviatoric tensor is deviatoric, and with respect to a
spherical tensor is spherical, the first term in the integrand can be
reformulated as
 \begin{align}
\tr \93{ \qqBde\dev{\02{\sym{\01{\qqAv\otimes\nablal}}}} } ,
 \end{align}
and the second term can be treated analogously. Next, it is easy to show
that the third term is \mm{\qqBv\cdot\01{\qqAde\cdot\nablal}}. Further,
in the terms that contain \m { \kdel{ij} }, the \m{j} index can be
changed to \m{i}, hence, in these terms we find the gradient of the
trace of a tensor. Therefore, the fourth term contains trace of a
deviatoric tensor---which is traceless---so this term gives zero
contribution. Finally, the fifth term contains a spherical tensor and
thus can be rewritten as
 \mm{ \f{1}{3}\pd_k\01{\qqAse}_{ii}\01{\qqBv}_k =
 \pd_k\01{\qqAse}_{ik}\01{\qqBv}_i }.
To sum up, we find 
 \begin{align}
\brmin{\qA}{\qB} & = -\int_\qV \f{1}{\qrho}
\Bigg[\qqBv\cdot\01{\qqAde\cdot\nablal + \qqAse\cdot\nablal} +
\tr\03{\qqBde\dev{\02{\sym{\01{\qqAv\otimes\nablal}}}}}
\mathrel+
 \nonumber  \\
 & \hskip 5em
+ \tr\03{\qqBse\sph{\02{\sym{\01{\qqAv\otimes\nablal}}}}}\Bigg]\dd\qV
= -\brmin{\qB}{\qA},
 \end{align}
antisymmetry is revealed.

Since \m{\qMmin} is independent of the state variables and is
antisymmetric, the generalized Poisson bracket also satisfies the Jacobi
identity \cite{ottinger:2005}.

Now let us turn towards the irreversible side: the operator matrix \m {
\qMpos } can be constructed from the second term
of the time evolution equation \re{timeevolution} and the degeneracy
condition \re{crit1}; we find
 \begin{align}  \label{irrop}
\qMpos=\begin{pmatrix}
\qM{11}&\zero\singledot&\zero\singledot&\qM{14}&\qM{15}&\qM{16} \\
\zero\dyad&0&0&\zero&0&0 \\ \zero\dyad&0&0&\zero&0&0\\
\qM{41}&\zero\doubledot&\zero\doubledot&\qM{44}&\qM{45}&\qM{46} \\
\qM{51}&0&0&\qM{54}&\qM{55}&0 \\ \qM{61}&0&0&\qM{64}&0&\qM{66}
\end{pmatrix}
 \end{align}
with\footnote{An outline and order of obtaining the components is as
follows:
\m { \qM{24} = 0 } and \m { \qM{34} = 0 } from \re{timeevolution},
\m { \qM{42} = 0 } and \m { \qM{43} = 0 } from \re{sym},
\m { \qM{56} = 0 } and \m { \qM{65} = 0 } from \re{timeevolution},
\m { \qM{44} } from \re{timeevolution},
\m { \qM{54} } and \m { \qM{64} } from \re{timeevolution},
\m { \qM{45} } and \m { \qM{46} } from \re{sym},
\m { \qM{41} } from \re{crit1},
\m { \qM{14} } from \re{sym},
\m { \qM{51} } and \m { \qM{55} } from \re{timeevolution},
\m { \qM{61} } and \m { \qM{66} } from \re{timeevolution},
\m { \qM{15} } and \m { \qM{16} } from \re{sym},
\m { \qM{11} } from \re{crit1},
the unconstrained remaining components can be set to zero.

}
 \begin{align}
\qM{11}&=-\f{\qld{11}}{\qrho^2}\03{\qT\dev{\02{\sym{\01{\bullet\otimes\nablal}}}}}\cdot\nablal-\f{\qls{11}}{\qrho^2}\03{\qT\sph{\02{\sym{\01{\bullet\otimes\nablal}}}}}\cdot\nablal,
 \nonumber  \\
\qM{14}&=\f{1}{\qrho^2}\03{\02{\qld{11}\dev{\02{\sym{\01{\qv\otimes\nablal}}}}+\qld{12}\01{-\qrho\qT\dxi}+\qls{11}\sph{\02{\sym{\01{\qv\otimes\nablal}}}}+\qls{12}\01{-\qrho\qT\sxi}}\bullet}\cdot\nablal,
 \nonumber  \\
\qM{15}&=
 \Mbox{
\f{\qld{12}}{\qrho}\01{\qT\bullet}\cdot\nablal,
 }
\qM{16}=\f{\qls{12}}{\qrho}\01{\qT\bullet}\cdot\nablal,
 \nonumber  \\
\qM{41}&=-\f{\qld{11}}{\qrho^2}\tr\03{\dev{\02{\sym{\01{\qv\otimes\nablal}}}}\dev{\02{\sym{\01{\bullet\otimes\nablal}}}}}+\f{\qld{21}}{\qrho}\tr\03{\qT\dxi\dev{\02{\sym{\01{\bullet\otimes\nablal}}}}}
 \mathrel-
 \nonumber  \\  \label{Melem}
 & \hskip 1.2em
-\f{\qls{11}}{\qrho^2}\tr\03{\sph{\02{\sym{\01{\qv\otimes\nablal}}}}\sph{\02{\sym{\01{\bullet\otimes\nablal}}}}}+\f{\qls{21}}{\qrho}\tr\03{\qT\sxi\sph{\02{\sym{\01{\bullet\otimes\nablal}}}}},
 \\
\qM{44} & =
\f{\qld{11}}{\qrho^2\qT} \tr\03{\dev{\02{\sym{\01{\qv\otimes\nablal}}}}\dev{\02{\sym{\01{\qv\otimes\nablal}}}}} -
\f{2 \qldsym{12}}{\qrho}
\tr\03{\dxi\dev{\02{\sym{\01{\qv\otimes\nablal}}}}}
+ \qld{22}\qT\tr\01{\dxi\dxi}
\mathrel+
 \nonumber  \\
 & \hskip 1.2em
+ \f{\qls{11}}{\qrho^2\qT} \tr\03{\dev{\02{\sym{\01{\qv\otimes\nablal}}}}\dev{\02{\sym{\01{\qv\otimes\nablal}}}}} -
\f{2 \qlssym{12}}{\qrho}
\tr\03{\sxi\sph{\02{\sym{\01{\qv\otimes\nablal}}}}}
 + \qls{22}\qT\tr\01{\sxi\sxi},
 \nonumber  \\
\qM{45}&=
 \Mbox{
\f{\qld{12}}{\qrho}\tr\03{\dev{\02{\sym{\01{\qv\otimes\nablal}}}}\bullet}-\qld{22}\qT\tr\01{\dxi\bullet},
 }
\qM{46}=\f{\qls{12}}{\qrho}\tr\03{\sph{\02{\sym{\01{\qv\otimes\nablal}}}}\bullet}-\qls{22}\qT\tr\01{\sxi\bullet},
 \nonumber  \\
\qM{51}&=
 \Mbox{
-\f{\qld{21}}{\qrho}\qT\dev{\02{\sym{\01{\bullet\otimes\nablal}}}},
 }
\qM{54}=
 \Mboxx{
\f{\qld{21}}{\qrho}\dev{\02{\sym{\01{\qv\otimes\nablal}}}}-\qld{22}\qT\dxi,
 }
\qM{55}=\qld{22}\qT,
 \nonumber  \\
\qM{61}&=
 \Mbox{
-\f{\qls{21}}{\qrho}\qT\sph{\02{\sym{\01{\bullet\otimes\nablal}}}},
 }
\qM{64}=
 \Mboxx{
\f{\qls{21}}{\qrho}\sph{\02{\sym{\01{\qv\otimes\nablal}}}}-\qls{22}\qT\sxi,
 }
\qM{66}=\qls{22}\qT.
 \nonumber
 \end{align}
One can notice that this \m{\qMpos} is not symmetric---see the elements
that contain \m{\qld{12}} vs.\ \m{\qld{21}}, as well as the ones with
\m{\qls{12}} vs.\ \m{\qls{21}}. As mentioned in Section~\ref{intsum},
when  we have no additional microscopic or experimental information
about \m { \qxi } and about the corresponding coefficients \m{\qld{}},
\m { \qls{} } then we cannot exclude that
antisymmetric parts \m{\qldskw{}}, \m{\qlsskw{}}
appear in the dynamics.

On the other side, positive semidefiniteness can be shown by
reformulating the integrand of the irreversible bracket \m{\brpos{A}{A}}
to a quadratic expression. More closely, we can form a matrix that
contains nonnegative elements and the Onsagerian coefficients, and
\re{requird}--\re{requirs} just prove to be the conditions that ensure
positive semidefiniteness. The calculation is straightforward but
lengthy.

Actually, the whole realization of \m { \qxi }-based rheology provided
above is straightforward (if lengthy), and is expected to work for
nonconstant coefficient matrices \m { \qld{} }, \m { \qls{} } as well.
However, specifically for constant coefficients, an alternative version
is also possible: implementing the antisymmetric part of the coefficient
matrices, that is, the constants \m { \qldskw{12} }, \m { \qlsskw{12} },
in the reversible part of the time evolution.

For this case, let us use the prepared \re{ons1sa}--\re{ons4sa}
form of the Onsagerian equations. Rearranging the time evolution equation
is straightforward, and we find
for the alternative reversible operator matrix \m { \QMmin }
 \begin{align}  \label{Lsymskw}
\QMmin=\begin{pmatrix}
0&\f{1}{\qrho}\bullet\cdot\nablal&\f{1}{\qrho}\bullet\cdot\nablal&
\zero&-\f{\qldskw{12}}{\qrho}\bullet\cdot\nablal&
-\f{\qlsskw{12}}{\qrho}\bullet\cdot\nablal\\
\f{1}{\qrho}\dev{\02{\sym{\01{\bullet\otimes\nablal}}}}&0&0&\zero&0&0\\
\f{1}{\qrho}\sph{\02{\sym{\01{\bullet\otimes\nablal}}}}&0&0&\zero&0&0\\
\zero\singledot&\zero\doubledot&\zero\doubledot&0&\zero\doubledot&\zero\doubledot\\
-\f{\qldskw{12}}{\qrho}\dev{\02{\sym{\01{\bullet\otimes\nablal}}}}&0&0&\zero&0&0\\
-\f{\qlsskw{12}}{\qrho}\sph{\02{\sym{\01{\bullet\otimes\nablal}}}}&0&0&\zero&0&0
\end{pmatrix} ,
 \end{align}
while the elements of the alternative irreversible operator matrix \m {
\QMpos } are very similar to \re{Melem}: we just have to change all
\m{\qld{12}} and \m{\qld{21}} to \m{\qldsym{12}} and, similarly,
\m{\qls{12}} and \m{\qls{21}} to \m{\qlssym{12}}.

The degeneracy criteria, antisymmetry of \m{\QMmin}, the Jacobi identity
for the generalized Poisson brackets, and positive semidefiniteness for
\m{\QMpos} prove to be satisfied. Moreover, in this case the symmetric
property of \m{\QMpos} also holds. 

We repeat that this alternative realization is valid only for constant
Onsagerian coefficients as otherwise the Jacobi identity were violated.

This latter variant appears rather counter-intuitive from principal
point of view but may be beneficial for numerical solutions, e.g., to
have as much symplectic part in the numerical scheme as possible---see
\cite{ottinger:2018,shang-ottinger:2018} about the importance of this.

\section{Temperature as state variable}

For mechanical engineering applications and evaluations of experiments
(see, \eg \cite{asszonyi-csatar-fulop:2016}), it can be beneficial to
use temperature, instead of entropy, as state variable. Then the
collection of state variables is
 \begin{align}
\label{statevarT}
\tx=\01{\qv,\deps,\seps,\qT,\dxi,\sxi}.
 \end{align}

To keep the discussion as concrete and basic as possible, let us choose
the simplest constitutive equation for the internal energy, linear in
temperature with constant specific heat \m{\qc}, containing elastic
energy related to Hooke's law,
 \begin{align}
\el{\qsig}=\dev{\young}\deps+\sph{\young}\seps ,
 \qquad
\dev{\young}=2\qG ,
 \quad
\sph{\young}=3\qK ,
 \end{align}
and neglecting thermal expansion---which is manifested in the separation of
strain dependence and temperature dependence---:
 \begin{align}  \label{eintT}
\intern{\te} \01{\tx} = \ther{\te} \01{\qT} + \ela{\te} \01{\deps,\seps} =
\qc\qT + \f{\dev{\young}}{2\qrho} \tr\91{{\deps\deps}} +
\f{\sph{\young}}{2\qrho} \tr\31{{\seps\seps}} .
 \end{align}
Temperature has the same relationship to specific entropy as seen in
Section~\ref{intgen}, now utilized in the opposite direction (\ie what
is a variable and what is a function): The
thermodynamical consistency condition
 \mm{\f{\dd \tqs}{\dd \qT} = \f{1}{\qT} \f{\dd \ther{\te}}{\dd \qT}}
that follows from the Gibbs relation [and which is the manifestation of
the third equation of \re{partialderivatives}]
leads to
 \begin{align}
\label{sofT}
\tqs \91{\tx} = \qsaux + \qc \ln \f{\qT}{\qT_\aux},
 \end{align}
with auxiliary constants \m{\qsaux}, \m{\qT_\aux}, and the extension
\re{extentropy} induces
 \begin{align}  \label{extentropyy}
\ts \9 1 { \tx } = \tqs \9 1 { \tx } -
\f{1}{2}\tr{\91{\dxi\dxi}}-\f{1}{2}\tr{\31{\sxi\sxi}} ,
 \end{align}
or, expressing temperature,
 \begin{align}  \label{Tofs}
\qT \91{ \tqs \21{\ts, \dxi, \sxi} } = \qT_\aux \exp
\91{ \f{\tqs \21{\ts,\dxi,\sxi}-\qsaux}{\qc} } .
 \end{align}

Now the energy and entropy functionals are 
 \begin{align}
\qET & = \int_\qV \qrho \tot{\te} \dd\qV =
\int_\qV \qrho \92{ \f{1}{2} \qv \cdot \qv + \qc\qT +
\f{\dev{\young}}{2\qrho} \tr \91{ {\deps\deps} } +
\f{\sph{\young}}{2\qrho} \tr \31{ {\seps\seps}} } \dd \qV ,
 \\
\qST & = \int_\qV \qrho \ts \dd \qV =
\int_\qV \qrho \92{ \qsaux + \qc \ln \f{\qT}{\qT_\aux} - \f{1}{2}
\tr \91{ \dxi\dxi } - \f{1}{2} \tr { \31{\sxi\sxi} } } \dd \qV ,
 \end{align}
with corresponding functional derivatives
 \begin{align}
\f{\var \qET}{\var \tx}=
\begin{pmatrix}
\qrho\qv\\ \dev{\young}\deps\\ \sph{\young}\seps\\ \qrho\qc\\ \zero\\ \zero
\end{pmatrix},
 \qquad\qquad
\f{\var \qST}{\var \tx}=
\begin{pmatrix}
\zero\\ \zero\\ \zero\\ \f{\qrho\qc}{\qT}\\ -\qrho\dxi\\ -\qrho\sxi
\end{pmatrix}.
 \end{align}

We perform a transformation of variables \mm { \qx \to \tx }, to which
the transformation (operator) matrix
 \mm { \qQ = \f{\var \tx}{\var \qx} }
is accompanied. This \m { \qQ } can be used to establish the
relationship between the original and transformed reversible and
irreversible operator matrices \cite{ottinger:2005}:
 \begin{align}
\label{vartran}
\tMmin=\qQ\qMmin\qQ^\Trans,\qquad\qquad\tMpos=\qQ\qMpos\qQ^\Trans .
 \end{align}
In the present current special case, we change only the fourth
state variable (from \m{\hs} to \m{\qT}), so only the fourth row of
\m{\qQ} contains nontrivial elements. Furthermore, since \re{Tofs} does
not contain nonlocal (gradient) terms, we can realize the transformation
directly in the form
 \begin{align}
\qQ=\f{\pd \tx}{\pd \qx}=
\begin{pmatrix}
\ident&\zero&\zero&\zero&\zero&\zero\\
\zero&\identf&\zero&\zero&\zero&\zero\\
\zero&\zero&\identf&\zero&\zero&\zero\\
\zero&\zero&\zero&\f{1}{\qc}\qT&\f{1}{\qc}\qT\dxi&\f{1}{\qc}\qT\sxi\\
\zero&\zero&\zero&\zero&\identf&\zero\\
\zero&\zero&\zero&\zero&\zero&\identf
\end{pmatrix},
 \end{align}
where \m{\identf} denotes the fourth order identity tensor [the identity
that maps tensors to tensors, \ie themselves]. Then, using \re{vartran}
yields\footnote{Results \re{LLT}, \re{MelemT} can also be derived
directly from the time evolution formula and the degeneracy conditions.}
 \begin{align}  \label{LLT}
\tMmin=\qMmin
 \end{align}
so all the requirements of GENERIC---antisymmetry, Jacobi identity,
degeneracy---prove to hold for \m { \tMmin }, and we find that the
structure of \m{\tMpos} is the same as of \m{\qMpos} [see \re{irrop}],
with the elements
 \begin{align}
\tM{11}&=-\f{\qld{11}}{\qrho^2}\03{\qT\dev{\02{\sym{\01{\bullet\otimes\nablal}}}}}\cdot\nablal-\f{\qls{11}}{\qrho^2}\03{\qT\sph{\02{\sym{\01{\bullet\otimes\nablal}}}}}\cdot\nablal,
 \nonumber  \\
\tM{14}&=\f{1}{\qrho^2\qc}\03{\qld{11}
\dev{\02{\sym{\01{\qv\otimes\nablal}}}}\qT \bullet
\mathrel +
\qls{11}\sph{\02{\sym{\01{\qv\otimes\nablal}}}}\qT\bullet}\cdot\nablal,
 \nonumber  \\
\tM{15}&=
 \Mboxxx{
\f{\qld{12}}{\qrho}\01{\qT\bullet}\cdot\nablal,
 }
\tM{16}=\f{\qls{12}}{\qrho}\01{\qT\bullet}\cdot\nablal,
 \nonumber  \\
\tM{41}&=-\f{\qT}{\qrho^2\qc}\tr\03{\qld{11}\dev{\02{\sym{\01{\qv\otimes\nablal}}}}\dev{\02{\sym{\01{\bullet\otimes\nablal}}}}+\qls{11}\sph{\02{\sym{\01{\qv\otimes\nablal}}}}\sph{\02{\sym{\01{\bullet\otimes\nablal}}}}},
 \nonumber  \\  \label{MelemT}
\tM{44}&=
\f{\qT}{\qrho^2\qc^2}
\tr\03{\qld{11}\dev{\02{\sym{\01{\qv\otimes\nablal}}}}
\dev{\02{\sym{\01{\qv\otimes\nablal}}}} \bullet
 \mathrel+
\qls{11}\sph{\02{\sym{\01{\qv\otimes\nablal}}}}\sph{\02{\sym{\01{\qv\otimes\nablal}}}}\bullet},
 \\
\tM{45}&=
 \Mboxxx{
\f{\qld{12}}{\qrho\qc}\qT\tr\03{\dev{\02{\sym{\01{\qv\otimes\nablal}}}}\bullet},
 }
\tM{46}=\f{\qls{12}}{\qrho\qc}\qT\tr\03{\sph{\02{\sym{\01{\qv\otimes\nablal}}}}\bullet},
 \nonumber  \\
\tM{51}&=
 \Mboxxx{
-\f{\qld{21}}{\qrho}\qT\dev{\02{\sym{\01{\bullet\otimes\nablal}}}},
 }
\tM{54}=
 \Mboxxxx{
\f{\qld{21}}{\qrho\qc}\qT\dev{\02{\sym{\01{\qv\otimes\nablal}}}},
 }
\tM{55}=\qld{22}\qT,
 \nonumber  \\
\tM{61}&=
 \Mboxxx{
-\f{\qls{21}}{\qrho}\qT\sph{\02{\sym{\01{\bullet\otimes\nablal}}}},
 }
\tM{64}=
 \Mboxxxx{
\f{\qls{21}}{\qrho\qc}\qT\sph{\02{\sym{\01{\qv\otimes\nablal}}}},
 }
\tM{66}=\qls{22}\qT.
 \nonumber
 \end{align}
The variable transformation is expected to preserve the structure of
GENERIC (\cite{ottinger:2005}, page~26 in Section~1.2.4).
Indeed, by
substituting \re{eintT} into \re{energybalance} and rewriting it in
terms of temperature, the evolution equation for \m{\qT} is obtained,
and turns out to coincide with the fourth row of
\m{\tMpos\f{\var\qST}{\var\tx}} so the whole evolution equation has been
preserved under the transformation.
 Meanwhile,
as in the variable \m { \hs } case \m { \qMpos } has proved
nonsymmetric for nonzero
 \m { \qldskw{} } or \m { \qlsskw{} },
\m { \tMpos } behaves the same way.

Now let us repeat moving the
 \m { \qldskw{} } and \m { \qlsskw{}}
related part of the dynamics to the reversible part.
We find the antisymmetric
 \begin{align}  \label{hLsymskw}
\TMmin=\begin{pmatrix}
0&\f{1}{\qrho}\bullet\cdot\nablal&\f{1}{\qrho}\bullet\cdot\nablal&
\TL{14}&-\f{\qldskw{12}}{\qrho}\bullet\cdot\nablal&
-\f{\qlsskw{12}}{\qrho}\bullet\cdot\nablal
 \\
\f{1}{\qrho}\dev{\02{\sym{\01{\bullet\otimes\nablal}}}}&0&0&\zero&0&0\\
\f{1}{\qrho}\sph{\02{\sym{\01{\bullet\otimes\nablal}}}}&0&0&\zero&0&0\\
\TL{41}&\zero\doubledot&\zero\doubledot&0&\zero\doubledot&\zero\doubledot\\
-\f{\qldskw{12}}{\qrho}\dev{\02{\sym{\01{\bullet\otimes\nablal}}}}&0&0&\zero&0&0\\
-\f{\qlsskw{12}}{\qrho}\sph{\02{\sym{\01{\bullet\otimes\nablal}}}}&0&0&\zero&0&0
\end{pmatrix}
 \end{align}
with
 \begin{align}
\nonumber
\TL{14}&=-\f{1}{\qrho\qc}\01{\qldskw{12}\qT\dxi\bullet
\mathrel +
\qlsskw{12}\qT\sxi\bullet}\cdot\nablal,\\
\TL{41}&=-\f{\qT}{\qrho\qc}\tr\93{\qldskw{12}\dxi
\dev{\92{\sym{\91{\bullet\otimes\nablal}}}}+
\qlsskw{12}\sxi\sph{\92{\sym{\91{\bullet\otimes\nablal}}}}},
 \end{align}
and that \m{\TMpos} can be obtained from \m { \tMpos } like \m { \QMpos
} from \m { \qMpos }, \ie  changing all \m{\qld{12}} and \m{\qld{21}} to
\m{\qldsym{12}} and \m{\qls{12}} and \m{\qls{21}} to \m{\qlssym{12}}.
Symmetricity and positive semidefiniteness of \m{\tMpos}, the degeneracy
criteria, as well as the Jacobi identity related to \m { \TMmin } are
all satisfied.

It is to be noted here that, while the Jacobi property of \m { \TMmin }
is
foreseen
on general grounds -- any variable transformation
is expected to
preserve the Jacobi identity (\cite{ottinger:2005}, page~26
in Section~1.2.4) --, checking it directly is
nonstraightforward. The difficulty is related to the task of identifying
total divergences of multiple products among the numerous terms. When we
used the application \texttt{jacobi.m}
\cite{kroger-hutter-ottinger:2001} -- with appropriately increased
memory limit and run-time limit --, it could not confirm the Jacobi
identity (while it
 found its validity
for \m { \qMmin }, \m { \QMmin } and \m { \tMmin } seamlessly). Instead,
we have verified the Jacobi property of \m { \TMmin } both by hand and
via an own symbolic code. A key element was an advantageous convention
for classifying and grouping terms, which has reduced the number of
terms from thousands to hundreds, enabling to observe the remaining
cancellations.

\section{Conclusions}

The results can be summarized according to Table~\ref{tablazat}.

\begin{table}[!ht] \centering
 \begin{tabular}{lccccc}  \toprule
& \multicolumn{2}{c}{
variable
\m { \hs }
}
& & \multicolumn{2}{c}{
variable
\m { \qT }
}
 \\  \cmidrule{2-3} \cmidrule{5-6}
& \subheader & & \subheader
 \\  \midrule
\m { \qMmin } fulfils Jacobi & \YES & \YES & & \YES & \YES
 \\
\m { \qMpos } is symmetric & \NO\ if \m {
\qldskw{} , \qlsskw{}
\ne 0 }
& \YES & & \NO\ if \m {
\qldskw{} , \qlsskw{}
\ne 0 } & \YES
 \\  \bottomrule
 \end{tabular}
\caption{How the four versions behave with respect to generic GENERIC
expectations.}
\label{tablazat}
\end{table}

As a task for the future, a finite deformation version would be welcome.
How deeply this will require to address objectivity and spacetime
compatibility aspects \cite{fulop-van:2012,fulop:2015} is an open
question.

In parallel, the current small-strain version could be numerically (\eg
along the lines of \cite{ottinger:2018,shang-ottinger:2018}) applied for
concrete problems. For example, the recent analytical results
in \cite{fulop-szucs:2018} promise methodologically useful
outcome since those considerations are done in the force equilibrial
approximation [zero lhs in \re{mombalancee}, an approximation valuable
for various engineering situations], which is a challenge for
GENERIC with its explicite time evolution formulation. Principal as well
as numerically working solutions to this compelling situation can
provide fruitful contribution to both science and engineering.

\section*{Acknowledgement}
The authors thank Markus H\"utter for his assistance concerning the
initial steps, Hans Christian \"Ottinger, Miroslav Grmela, and Michal
Pavelka for valuable remarks, the participants of the workshop IWNET
2018 for the useful discussions, and Karl Heinz Hoffmann and
the Reviewers for helpful suggestions.

The work was supported by the Hungarian grants National Research,
Development and Innovation Office -- NKFIH 116375 and NKFIH KH 130378,
and by the Higher Education Excellence Program of the Ministry of Human
Capacities in the frame of Nanotechnology research area of Budapest
University of Technology and Economics (BME FIKP-NANO).

\end{document}